\documentclass{article}

\usepackage{graphicx}

\def\be{\begin{equation}}
\def\ee#1{\label{#1}\end{equation}}
\title{Inflationary and dark energy regimes \\
in  2+1 dimensions}
\author{M. H. Christmann, F. P. Devecchi, G. M. Kremer and C. M. Zanetti\\
Departamento de F\'\i sica, Universidade Federal do Paran\'a\\
Caixa Postal 19044, 81531-990, Curitiba, Brazil
}

\begin{document}

\maketitle

\begin {abstract}  
In this work we investigate the behavior of three-dimensional (3D)
 cosmological
models. The simulation
 of
inflationary  and dark-energy-dominated  eras are among the possible results
in these  3D formulations; taking as starting point the results obtained
 by Cornish and Frankel.
 Motivated  by those results, we  investigate, first,  the inflationary 
case where we consider a two-constituent cosmological fluid: the
scalar field represents the hypothetical inflaton which is in gravitational
interaction with a matter/radiation contribution.
For the description of an old universe,  it is
possible to simulate its evolution  starting with a matter dominated universe 
that faces a decelerated/accelerated transition due to the presence of the 
additional constituent (simulated by the scalar field or ruled by an exotic 
equation of state) that plays the role of dark energy.  
We obtain, through  numerical analysis, the evolution in time of 
the scale
factor, the acceleration, the energy densities, and the hydrostatic
pressure of the constituents. 
The alternative scalar cosmology proposed by Cornish and Frankel is also under
investigation in this work. In this case an  inflationary model 
can be constructed when another 
non-polytropic
equation of state ( the van der Waals equation) is used to simulate the 
behavior of an early 3D universe.
\end{abstract}

\section {Introduction}

The investigation of cosmological models  in lower dimensions 
 provide 
 technical insights that can  be applied to ``realistic''  models\cite{Brown}. 
In the case of 
2+1 dimensions (3D models) several analysis were done mainly on the 
simulation of matter/radiation eras\cite{Cornish,Giddings}. 2D universes
were also considered in the literature\cite{Deve}: here, besides 
the matter/radiation periods, it was possible to simulate inflationary eras.
In  the case of homogeneous and isotropic models  several alternative 
formulations were tested like the Jackiw-Teitelboim formulation\cite{Deve} 
in 2D and the
scalar gravity model in 3D\cite{Cornish}. In fact, in the 2D case the Einstein
field equations are void of information and therefore the search for 
an alternative theory is mandatory\cite{Brown}. 

One remarkable fact in 3D Einsteinian models is that the Riemann curvature 
tensor is 
zero outside 
sources, so there is no free propagation of 
the gravitational field in these
formulations. As a consequence, the theories do not possess a Newtonian limit
\cite{Cornish,Giddings}. However, in a cosmological context 
the 3D space-time is supposed 
to be filled by the sources and in several cases a ``regular'' behavior 
follows. The basic ingredients in these models are 
the gravitational field equations
(where the Einstein equations or alternative dynamics\cite{Brown} 
are considered) the energy-momentum conservation law, the equations of state
(in a macroscopic description) for the sources and their correspondent 
field equations, in a curved-space-time description.

In this work we extend the analysis done in \cite{Cornish,Giddings} 
considering the possibility of inflationary and dark energy regimes in 3D 
cosmologies. We make use of thermodynamics of relativistic gases  and  
consider a two-constituents model with matter and a scalar field that is 
going to simulate the presence of the inflaton for a young universe
or dark energy for an old universe.
 The  simulation of interaction (through the gravitational field)
between the constituents is done using a dynamical pressure term in the 
energy momentum-tensor of the sources. We obtain the evolution in time 
of the scale factor, the energy densities of the constituents and  we 
focus mainly on the 
behavior of the acceleration of the expansion,  as a fundamental ingredient
to the classification of the cosmological regimes.

As we mentioned above another possibility suggested  in the literature      
was the analysis of a 3D scalar gravity
model\cite{Cornish}. Here we extend those results
 by using   an exotic equation of state (the van der Waals equation (vdW), 
proposed for 4D cosmological models in \cite{Capo1} )  
to simulate a three-eras regime that would correspond to 
an initial  inflationary period, followed by a decelerated era where matter 
or radiation would dominate and finally an accelerated period dominated 
by dark energy represented by a cosmological constant.

The article is structured as follows. 
In section II we make a brief panorama of the physical principles involved 
in 3D cosmological models and  in section III we present the analysis of a 
3D inflationary period. In section IV we consider an old universe where the 
dark energy is represented by an scalar field  or by the Chaplygin equation.
In section V we show the analysis of the 3D scalar model. Finally in section
VI we display our conclusions.

\section{Field equations  and conservation laws}

  In this section we make a brief review of the dynamics that rule 
the gravitational field and the sources in  3D space-times (for  a  
detailed presentation see for instance \cite{Brown,Cornish,Giddings}). One 
essential feature that appears in 3D is that the Riemann tensor components 
are zero outside sources \cite{Cornish,Giddings}. However,
  when we work in a cosmological context 
(taking a  universe filled by  sources), 
the 3D Einstein field equations permit the propagation 
of the gravitational field\cite{Cornish}:

  \begin{equation}
    R_{\mu\nu}-\frac{1}{2}\,R\,g_{\mu\nu}\,= -\kappa T_{\mu\nu} =
\,-2\pi\,G\,T_{\mu\nu}\,\, ,
  \end{equation}

  \noindent 
 where $G$ is the ``renormalized'' gravitational constant for 3D 
space-time \cite{Cornish}. As in the  4D case, the hypotheses of 
isotropy and homogeneity are
 represented
 by the  Robertson-Walker metric:

  \begin{equation}
    ds^2\,=\,dt^2-a(t)^2\left[dr^2+r^2d\theta^2\right]\,\,\label{1}\,\, ,    
  \end{equation}

\noindent where  $a(t)$ is the scale factor. The sources of the 
gravitational field  are seen
as an out-of-equilibrium composite fluid. The correspondent 
energy-momentum tensor is given by

\begin{equation}
    T^{\mu\nu}\,=\,(\sum _i[\rho _i+p_i]+\varpi)\,U^{\mu}U^{\nu}
-g^{\mu\nu}(\sum _i p_i+\varpi)\, ,
  \end{equation}
  
\noindent where  $\rho _i$ and $p_i$ ($i=1,2,...N$)
 are the energy density and the hydrostatic pressure of the sources,
 respectively. $\varpi$ is the dynamical pressure, related to the
viscosity of the cosmological  fluid.
When we work in a cosmological context and include
 the dynamical pressure it means that we are taking into account,
phenomenologically, the
interaction  between the constituents  through the
gravitational field \cite{dk}.
  The energy-momentum  tensor satisfies the correspondent  conservation  law
  $T^{\mu\nu}_{\,\,;\nu}=0$, that in a Co-moving frame reads

  \begin{equation}
\label{3}
   \sum _i \dot{\rho _i}+2\frac{\dot{a}}{a}\lbrace \sum _i [\rho _i+p_i]
+\varpi \rbrace =0.
  \end{equation}
  
  Solving the  Einstein equations  for the  3D  Robertson-Walker
  metric  we get
  \begin{equation}
    H^2\,= \kappa  \sum _i \rho _i
  \,\,\,\,\, \, , \,\,\,
    \frac{\ddot{a}}{a}\,=\,- \kappa (\sum _i p_i+\varpi).
  \end{equation}

\noindent   where $H=\dot{a}/a$ is the  Hubble function. In another possible 
case
(such in the case of the dark energy) the sources can be modelled as a 3D
scalar field ($\phi$). This means that the curved space-time  Klein-Gordon
equation is 
necessary. In an
isotropic and homogeneous  3D space-time it is given by

\begin{equation}
\ddot  \phi + 2 H \dot \phi = - V'(\phi)\, 
\end{equation}

\noindent where $V$ is the potential and  the exclusive dependence in time 
of $\phi$ is forced by the Robertson-Walker hypothesis.

Besides the 3D Einstein model other 3D gravitational theories has been 
proposed 
in the 
literature (such as the Einstein-Weyl model that includes 
torsion\cite{Brown,Jackiw,Giddings}). 
In \cite{Cornish} was considered the possibility of an scalar 3D cosmology
ruled by

\begin {equation}
R = -2\kappa T\, \, .
\end{equation}

\noindent    
This law gives in principle insufficient dynamical 
information for the 
gravitational field tensor. However, 
in Robertson-Walker cosmologies  there is only one field to 
be determined (the scale factor $a(t)$)and this makes possible to consider 
a model 
based on that scalar equation. In fact, in \cite{Cornish} a radiation 
dominated and a  matter dominated 3D universe were simulated using the scalar
cosmology. Another  property of this model is that it  permits a 
Newtonian limit
\cite{Cornish}.

  In  the following section    we consider first the 
3D Einstein theory 
to investigate the behavior of a young universe filled by a scalar field
(representing the inflaton)
and a matter constituent; both interacting through the gravitational field.

{
\section{ Inflationary Universe  in 3D}}

In this section we consider an inflationary universe. The main point here is 
to
investigate how the 3D Einstein model describes the  transition between an 
inflaton-dominated regime and the
beginning of a matter era; focusing on the behavior of the
  acceleration ($\ddot a$) 
and on the energy densities ($\rho _m$ , $\rho _{\phi }$). 
The cosmological fluid
is composed by two constituents: matter is represented by a barotropic
equation of state and the inflaton is an scalar field $\phi (t)$. As it 
was explained in the last section  the 
field equations and 
the energy-momentum conservation law  furnish
the dynamical relations between the  observables. 
Taking the particular case of  the
energy conservation law it  can be separated into two independent relations;
here we are using the fact that the scalar field behavior is ruled by
the  curved space-time
 Klein-Gordon 
equation. The system of equations can be written in the following form

\begin{equation}
    H^2= \kappa (\rho_m+\rho_{\phi}),
 \,\,\, \dot{H}+H^2=- \kappa (p_m+p_{\phi}+\varpi),
  \end{equation}

\begin{equation}
\label{}
    \dot{\rho}_m+2H(\rho_m+p_m+
\varpi)=0\, , \, 
    \dot{\rho}_{\phi}+2H(\rho_{\phi}+p_{\phi}
)=0\,.
  \end{equation}

\noindent The thermodynamical behavior of  both constituents is supposed to
be ruled by\cite{Peebles}

  \begin{equation}
    p_m\,=\,(\gamma-1)\rho_m \, , \,\,\,\,\, 
  p_{\phi}\,=\,(\nu-1)\rho_{\phi}\, \, ,
  \end{equation}
  
\begin{equation}
    \varpi\,=\,-\alpha(\rho_m+\rho_{\phi})\Theta\, , \quad
    (\Theta=\nabla^{\mu}U_{\mu}=2H)\, , 
  \end{equation}

\noindent where $\alpha$ is a viscosity parameter and $\varpi$ is the dynamical
pressure\cite{dk}. 
The combination of these expressions leads us to the following
 system  
  
       \begin{equation}
   \frac{\rho_m}{\rho_{\phi}^0}\,
=\,\left(1+\frac{\rho_m^0}{\rho_{\phi}^0}\right)\,H^2-
\left(\frac{1}{a}\right)^{2\nu}\, \,\,\, ,\,\,
  \frac{\rho_{\phi}}{\rho _{\phi}^0}\,=\,\left(\frac{1}{a}\right)^{2\nu}
  \end{equation}

  \begin{equation}
    \dot{H}\,=\,(2\alpha\,H-\gamma)\,H^2+
\frac{\gamma-\nu}{1+\rho_m^0/\rho^0_{\phi}}\left(\frac{1}{a}\right)^{2\nu}\,
,\end{equation}

  \noindent
  where  $\rho_m^0/\rho_{\phi}^0$ is the ratio  between the 
energy densities at $t=0$. For convenience, we use the normalization 
$ H^2\equiv
\frac{H^2}{H_0^2}$, $H_0^2 = \kappa (\rho _m^0 + \rho _{\phi}^0)$ 
and $ 2\alpha
\equiv \alpha H _ 0$.
The differential equation (13) is a non-linear, 
second order equation for the scale factor 
 $a(t)$; it  depends on four parameters 
   $\gamma$, $\nu$, $\alpha$ and the quotient
  $\rho_m^0/\rho_{\phi}^0$. 
To solve this equation we must specify those values 
and two initial conditions $a(0)$ and $\dot{a}(0)$ and  the intervals for the
barotropic  parameters.  These are shown in the following table

  \begin{center}
    \begin{tabular}{|c||c|}
      \hline
      {\small\it viscosity}  &$0<\alpha<1$ \\
      \hline
      {\small\it matter state equation } & $1<\gamma<2$ \\
      \hline
      {\small\it inflaton state equation } & $0<\nu<1$     \\
      \hline\hline
      $a(0)$         & $1.0$  \\
      \hline
      $\dot{a}(0)$   & $1.0$  \\
      \hline
   $r=\rho_m^0/\rho_{\phi}^0 $         & $ 0<r<0.5$  \\
\hline
    \end{tabular}
  \end{center}   

The numerical analysis of the system (11-12) give the following results 
for the evolution of the
physical quantities:

  \begin{figure}\vskip0.0truecm\begin{center}
\includegraphics[width=5.0cm]{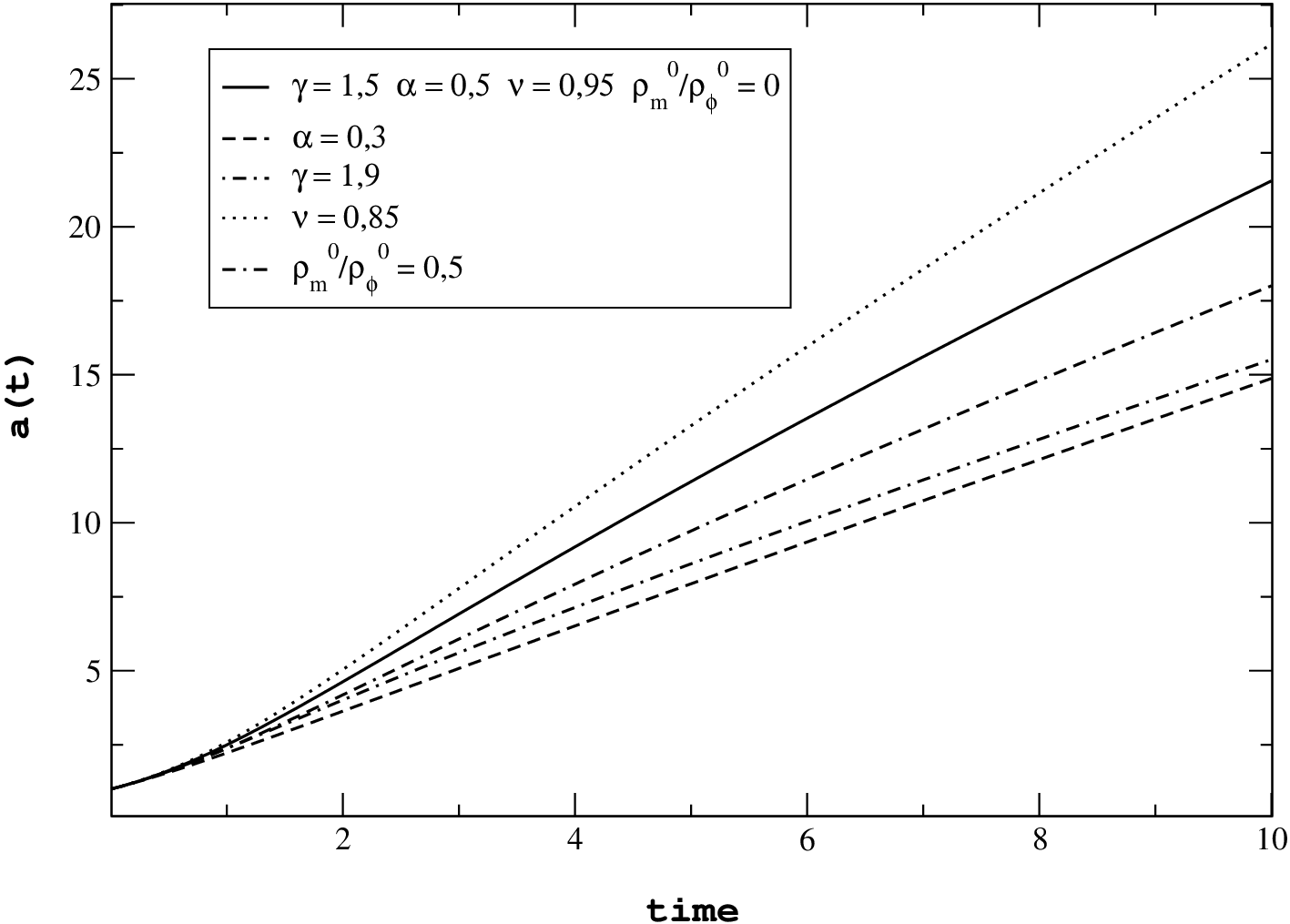}
\caption{evolution in time for scale factor a(t) taking different values of
the parameters}
\end{center}
\end{figure}

\noindent Figure 1 presents the behavior of the scale factor during the
period of inflation. The numerical results show that an increasing viscosity
 $\alpha$ implies into a faster expansion. On the other hand larger values
of $\gamma$ and $\nu$  (see the equations of state 10) furnish a slower 
expansion. Besides, for an increasing 
quotient $\rho ^0_m/\rho ^0_{\phi}$ the expansion is again slower.  

  \begin{figure}\vskip0.0truecm\begin{center}
\includegraphics[width=6.4cm]{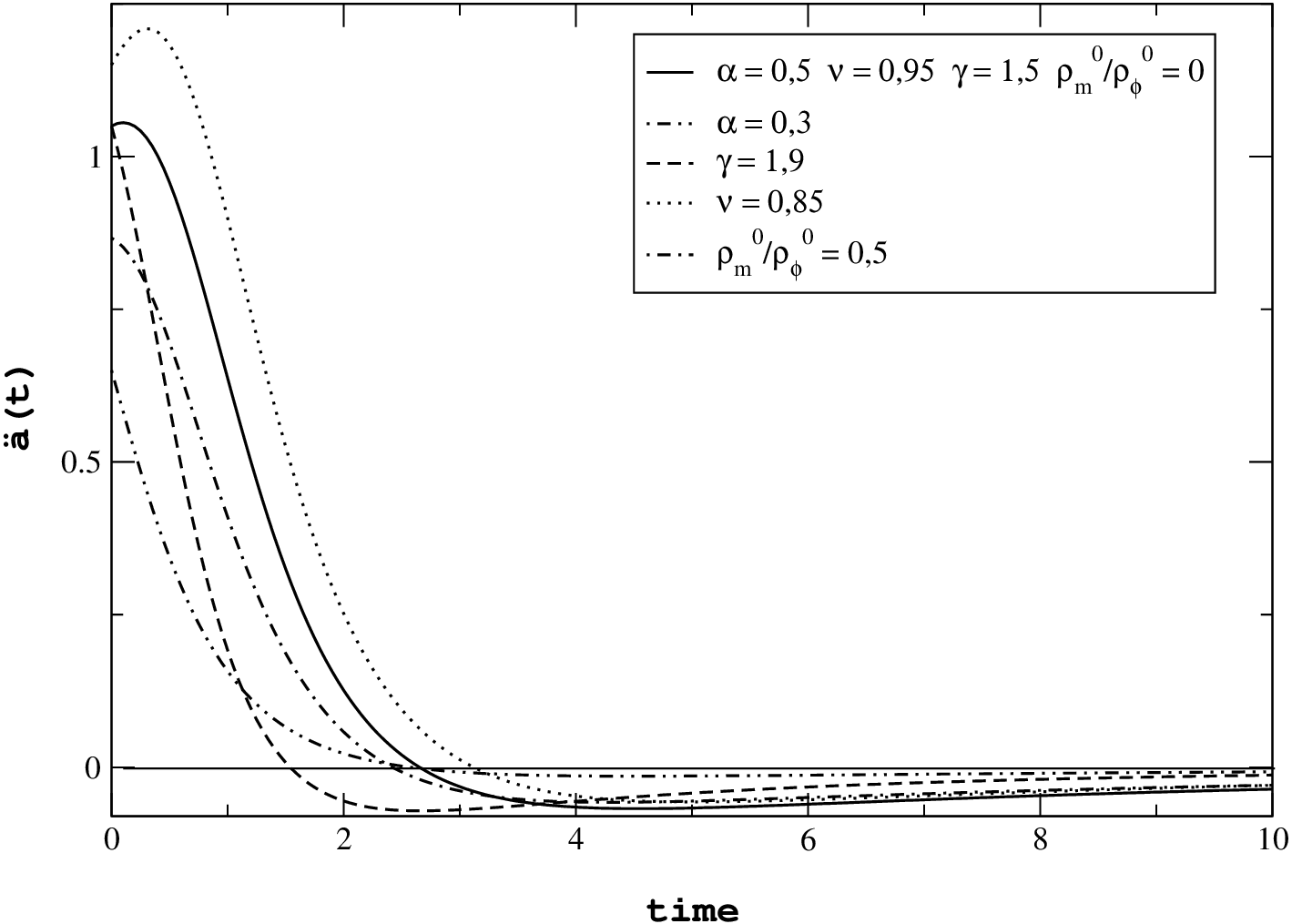}
\caption{behavior of acceleration during inflation for different sets of
parameters}
\end{center}
\end{figure}

\noindent In figure 2 we plot the behavior of acceleration function 
$\ddot a(t)$ for typical values of the parameters; the
numerical analysis shows the  transition of a strongly
accelerated, expanding universe  to a decelerated  regime representing the end
of the inflationary period and the beginning of a matter period. The results
show that  keeping the other parameters fixed the smaller the value of the
viscosity $\alpha$ the smoother is the transition. On the other hand if we
increase the ratio between the energy densities for $t=0$ we find again a
smoother transition to a decelerated period.

  \begin{figure}\vskip0.0truecm\begin{center}
\includegraphics[width=6.0cm]{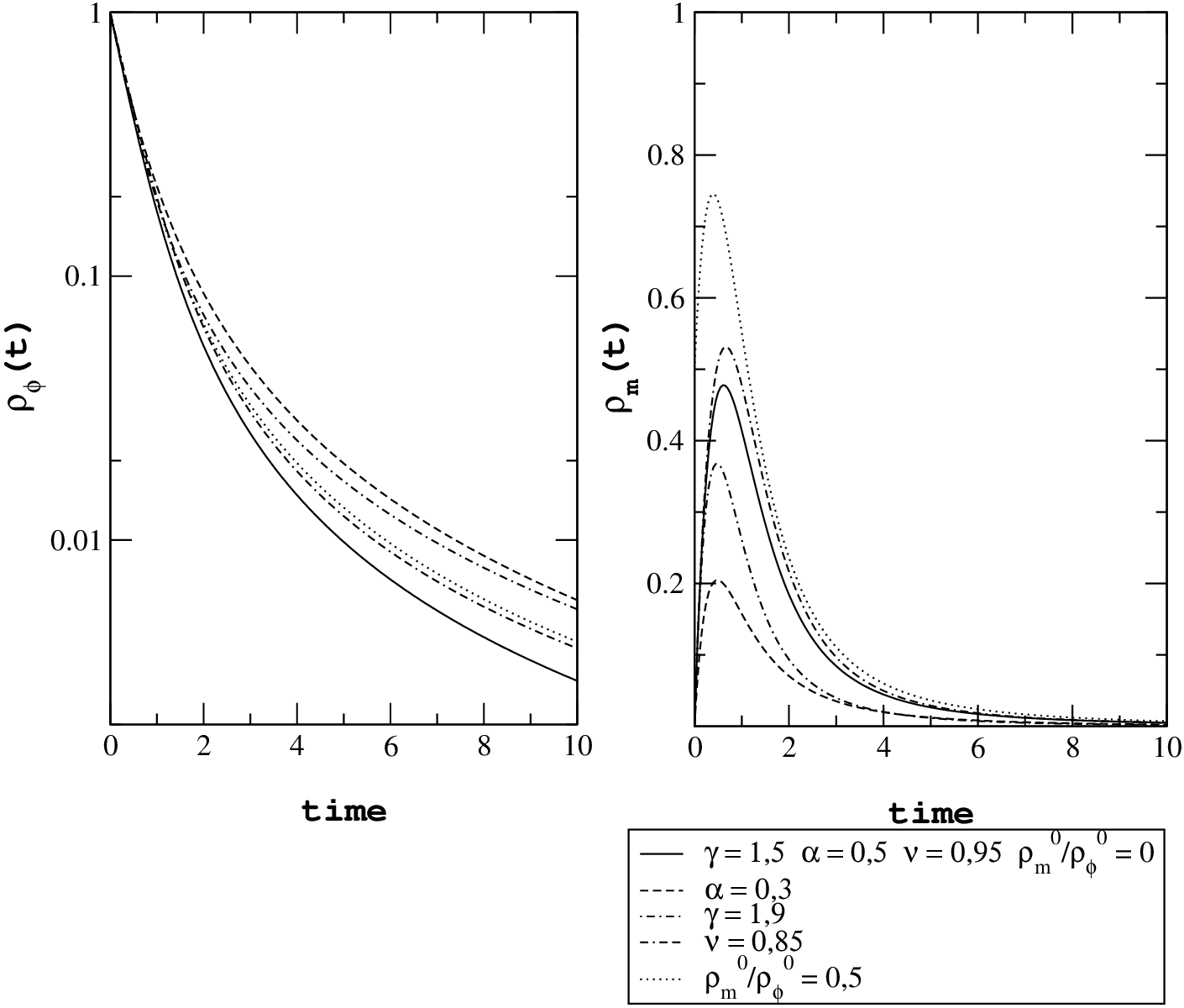}
\caption{evolution in time of energy densities during the expansion in the
inflationary regime}
\end{center}
\end{figure}

\noindent In figure 3 we show the  evolution in time of the energy densities. 
The
representation of  $\rho _{\phi}$ is in logarithmic scale. With the passing of
time matter starts
to predominate at the expense of the inflaton and gravitational field
energies; this fact is in accord with the transition to a decelerated era as
was shown in figure 2. 

It is worth mentioning  that  the  quotient $\rho ^0_m/\rho ^0_{\phi}$  is
essential to the classification of transitions:  for the inflationary
case we take a null (or small) value simulating that for $t=0$ we have a
clear predominance of inflaton energy. 
For the dark-energy-dominated case ( as we will see in the following section)
there is an opposite initial  situation, and this is fundamental to obtain a
transition between an initial decelerated period to a final accelerated era
(as it is desired).
These results show that the models give an adequate qualitative description
of the cosmological  regimes under analysis. Another important point is that
in the case of the inflationary scenario the gravitational interaction
between the constituents (represented by the dynamical pressure term)  is
relevant to the qualitative behavior of the acceleration.
 
On the other hand, the model also furnish solutions that do not describe
  transition regimes. For instance if the take the  usual values of figure
2 for the parameters and a viscosity  $\alpha \leq .29$ the transition does not
occur, and the model only provides an ever decelerated universe.

\vskip .5cm

{
\section{ Accelerated 3D universe dominated by dark energy }}

In this section we go forward in time  putting an old  3D universe 
under investigation. Two cases are analyzed: { i)} the dark energy 
constituent modeled as an scalar field (for the original 4D model see 
\cite{dk});  { ii)} the dark energy  is seen as 
a fluid ruled by the Chaplygin equation of state (proposed in 4D cosmological
  theories, 
see for instance \cite{Capo}).

In the first case, as in the
precedent 
section, the model considers  the cosmological fluid  consisting of a  
two-component gas with its  
energy-momentum 
tensor including
a dynamical pressure term ($\varpi$) to simulate the interaction between the
constituents via the gravitational field. The dynamics are again ruled by

  \begin{equation}
    \dot{H}\,=\,(2\alpha\, H-\gamma)H^2+ 
    \frac{\gamma-\nu}{1+\rho_m^0/\rho_X^0}\left(\frac{1}{a}\right)^{2\nu},
  \end{equation}

  \begin{equation} 
    \frac{\rho_m}{\rho _X^0}\,
=\,\left(1+\frac{\rho_m^0}{\rho_{X}^{0}}\right)H^2-
\left(\frac{1}{a}\right)^{2\nu},
 \,\,\, \frac{\rho_X}{\rho _X^0}\,=\,\left(\frac{1}{a}\right)^{2\nu}.
  \end{equation}

 \begin{figure}[h]
\vskip0.8truecm\begin{center}
\includegraphics[width=6.4cm]{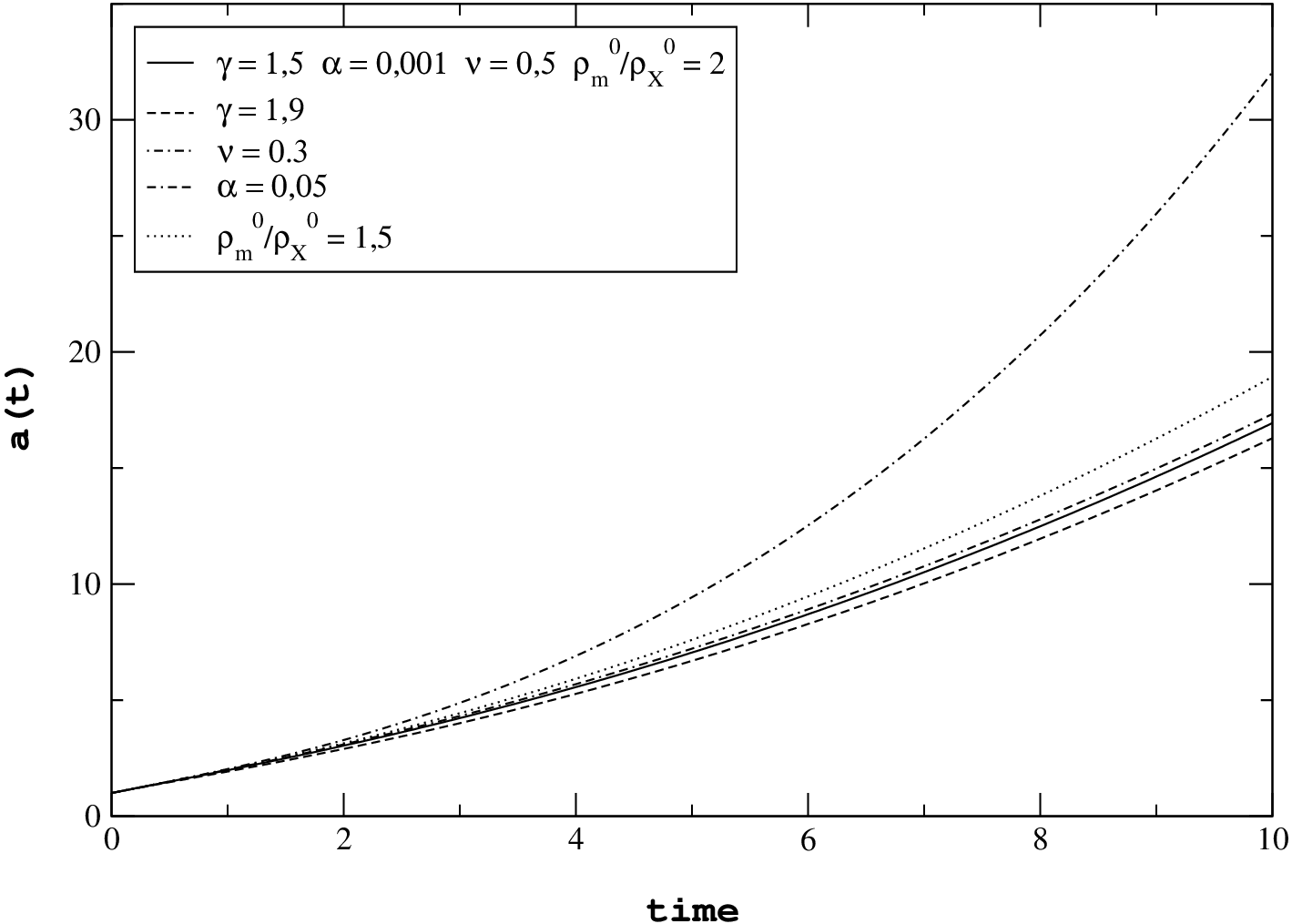}
\caption{evolution in time of scale factor in the
dark energy dominated regime}
\end{center}
\end{figure}

\noindent as was mentioned in the last section the quotient $\rho ^0_m/\rho
^0_{X}$ defines the regime and in this case we are taking initial values
that are compatible to the fact that matter dominates in the beginning of
the era. The initial conditions and clocks   are 
considered in a 
scale that is different from the previous section analysis; so the results are 
independent. As it is seen in figure 4 we still have a permanent expansion in
this old universe.
The results also show that an increasing initial ratio $\rho
^0_m/\rho ^0_{X}$ furnish a slower expansion. On the other hand only big
changes in the viscosity $\alpha$ give a substantial difference in the
expansion rate. For an increasing $\gamma $ we have an slower expansion in
this 3D old universe. 

 \begin{figure}\vskip0.8truecm\begin{center}
\includegraphics[width=6.4cm]{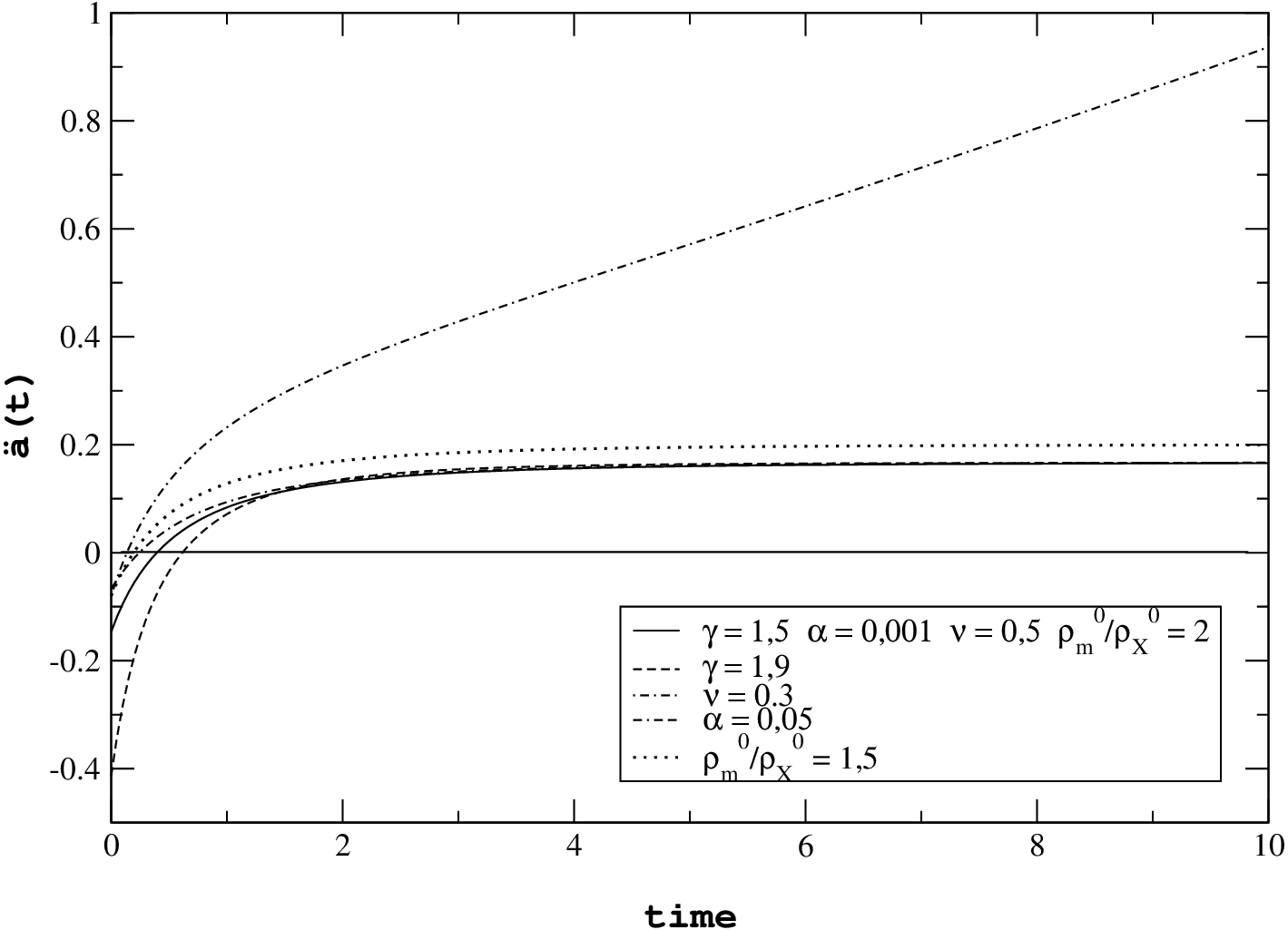}
\caption{evolution in time of acceleration  during the expansion in the
dark energy dominated regime}
\end{center}
\end{figure}

\noindent In figure 5 we plot the acceleration $\ddot a$ versus time; the 
results show that
an initial period of deceleration is followed by an accelerated era. The
extension of each period is related to the combination of parameters
$\alpha$, $\nu$, $\gamma$ and the quotient between the initial energy 
densities. The results show that there is a critical value  of
$\nu $ (around 0.3) that implies in a fast growing of the acceleration.
On the other hand, as it was expected, a larger quotient  $\rho
^0_m/\rho ^0_{X}$ gives an earlier transition to the accelerated regime, 
indicating that the dark energy starts dominating the cosmological fluid, 
opening an era of eternal accelerated expansion. 
This behavior is qualitatively similar to the 4D case\cite{Capo,dk}.

 \begin{figure}[h]\vskip0.8truecm\begin{center}
\includegraphics[width=7.6cm]{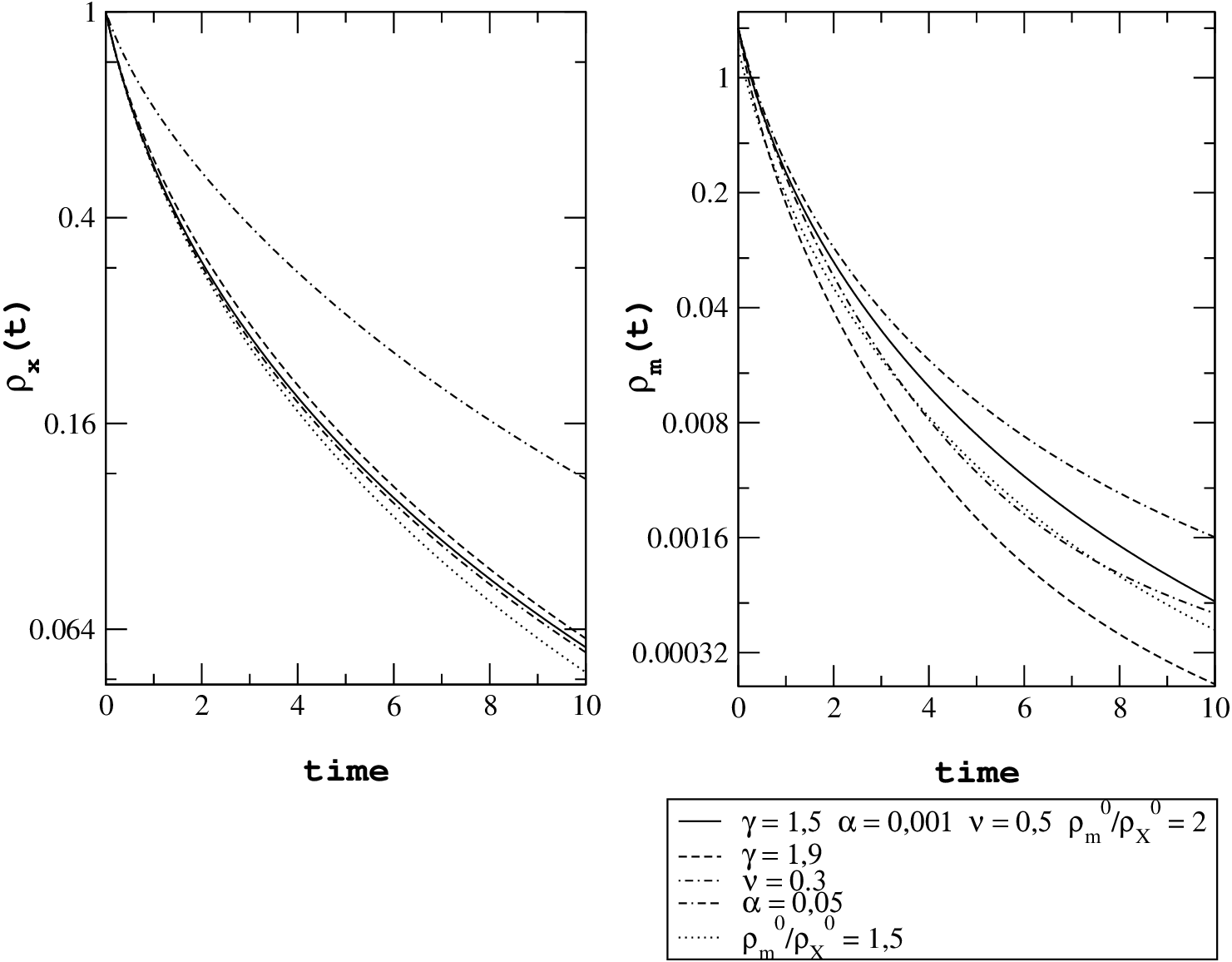}
\caption{evolution in time of energy densities during the expansion in the
dark energy dominated regime}
\end{center}
\end{figure}

\vskip .5cm

In the second case, the late 3D universe  is considered  as  
filled by a two-constituent fluid 
where matter is replaced
by  radiation  (with $p_r=\frac{1}{3}\rho _r$)
 and a equation of state known as the 
Chaplygin equation  to model the contributions of the dark
 energy (this was proposed for 4D models in \cite{Capo}). Again, the
interaction between the constituents is considered to occur via the 
gravitational field. The 
Chaplygin equation of state reads

\begin{equation}
 p_c = -{\frac{A}{\rho _c}} \,\, ,
\end{equation}

\noindent where $A$ is a positive parameter. This equation of state 
would correspond to the 
phenomenological modelling of 
a gas of membranes\cite{Capo}. The Einstein and conservation-law equations 
give in this case

\begin{eqnarray}
\dot{H}+\frac{3}{2}H^2=\frac{3+\frac{1}{a^4}}{2(1+\frac{\rho_r^0}
{\rho_c^0}\sqrt{1+\frac{1}{a_0^4}}\sqrt{1+\frac{1}{a^4}})}+2\alpha H^3, 
\label{sist}  
\end{eqnarray}

\begin{equation}
\frac{\rho_c}{\rho_c^0}
=\frac{\sqrt{1+\frac{1}{a^4}}}{\sqrt{1+\frac{1}{a_0^4}}},
\,\,\,
\frac{\rho_r}{\rho _r^0}
=\frac{\rho_c^0}{\rho_r^0}\left[\left(1+\frac{\rho_r^0}
{\rho_c^0}\right)H^2-\frac{\sqrt{1+\frac{1}{a^4}}}
{\sqrt{1+\frac{1}{a_0^4}}}\right].
\label{dr}
\end{equation}

\noindent The initial conditions in this case include $ a(0)=1, \dot a(t) = 1
, $ and ${{\rho _{c_0}} \over {\rho _{r_0}}} = 0.1$. 
 The results of the  numerical solutions of this system is 
shown in figure (7). 

\begin{figure}[h]\vskip0.3truecm
\begin{center}
\includegraphics[width=6.0cm]{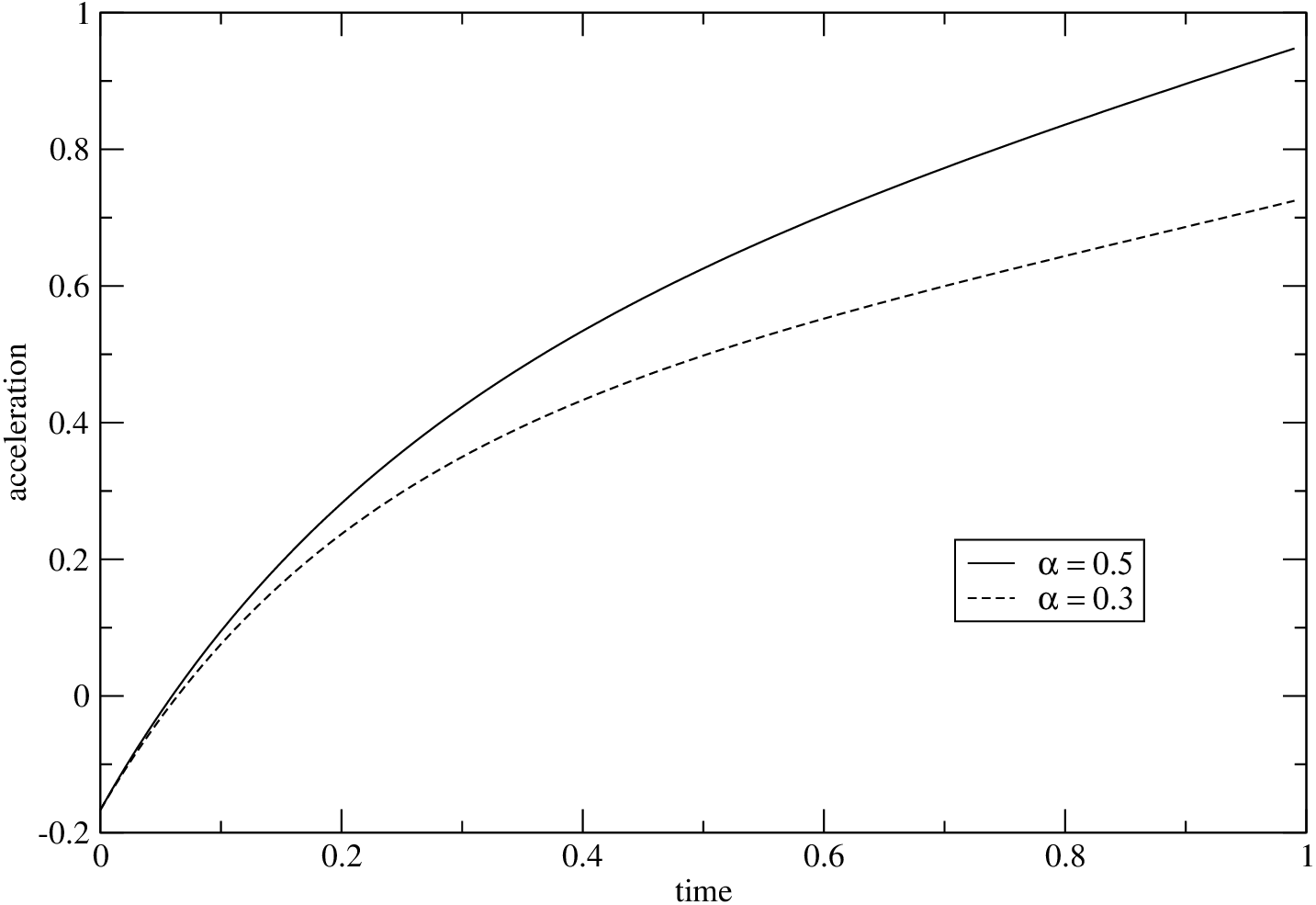}
\caption{acceleration with dark energy ruled by the Chaplygin 
equation, for different values of the viscosity parameter}
\end{center}
\end{figure}

This shows the acceleration behavior and is easily recognized as a transition
to positive values in later times. An increasing value of the viscosity 
parameter provide a more drastic transition to the accelerated regime. 
The energy densities behave in a similar way
as the scalar field case (see figure (6)). 
Putting $A=0$ the solutions  
confirm  that the Chaplygin gas is responsible for the positive accelerated 
regime. On the other hand the scale factor behavior correspond to a 
never ending expansion. Qualitatively, these results indicate a strong 
similarity with the ones showed in the last section: both the Chaplygin 
equation of
state (see figure (7)) and the scalar field formulation give a smooth 
transition to an 
accelerated regime, dominated by the dark energy constituent, 
coming from a precedent universe where matter/radiation were 
predominant.

 \vskip .2cm

\section{ Scalar gravity for  3D universes }

The verification of the singular features of 3D gravity embraces the
possibility  of testing other
models of gravitation in 3D space-times. Focusing on our purposes we mention
that  the  so-called scalar gravity 
model (``$R=T$'') 
was analyzed in a cosmological context in\cite{Cornish}. 
A remarkable feature of this model is that, contrary to the 3D Einstein 
case, its dynamics 
contains a Newtonian limit \cite{Cornish}. On the other hand, in principle, 
this model gives an incomplete information about the evolution of the 
gravitational field
\cite{Cornish}. However, when we impose the Robertson-Walker hypothesis
different cosmological scenarios can be consistently described; 
in fact several 
regimes 
involving matter  and/or radiation were considered in\cite{Cornish}.  
In this section we would 
like to extend those results 
investigating  the behavior of both  young   and  old universes  filled by
different combinations of constituents (including the inflaton and the 
dark energy)   ruled by a 3D scalar cosmology.

\begin{figure}[h]\vskip0.3truecm
\begin{center}
\includegraphics[height=3.2cm, width=5.0cm]{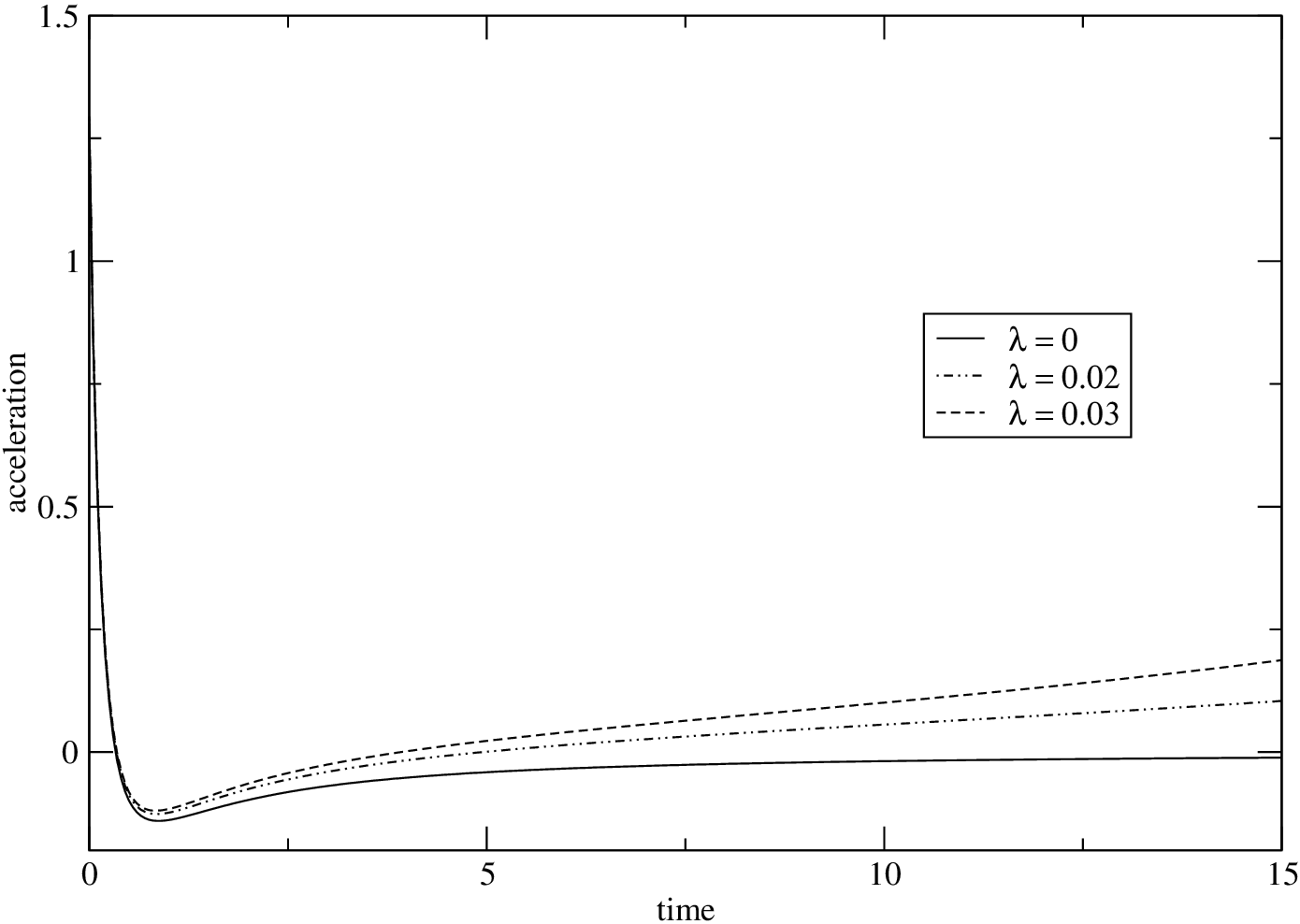}
\caption{acceleration for different values of cosmological 
constant $\lambda $}
\end{center}
\end{figure}

\noindent The  simplest case possible is to consider a one-constituent
 model
simulating, for example, radiation/matter
dominated periods.  This was done  in \cite{Cornish} with interesting
results. The dynamics is ruled by

\begin{equation}
2\frac{\ddot a}{a} = -(\frac{\dot a}{a})^2+\kappa (2p-\rho) + \lambda \,\, ,
\end{equation}

\noindent where $\lambda$ is a cosmological constant.
Our analysis focuses again  the possibility of describing transition
eras, in particular the end of an inflationary period. 
Inspired by the results obtained in 2D and 4D models 
we consider a one-constituent universe ruled by the van der
Waals equations of state \cite{Capo1}.

\begin{figure}[h]\vskip0.1truecm
\begin{center}
\includegraphics[width=6.6cm]{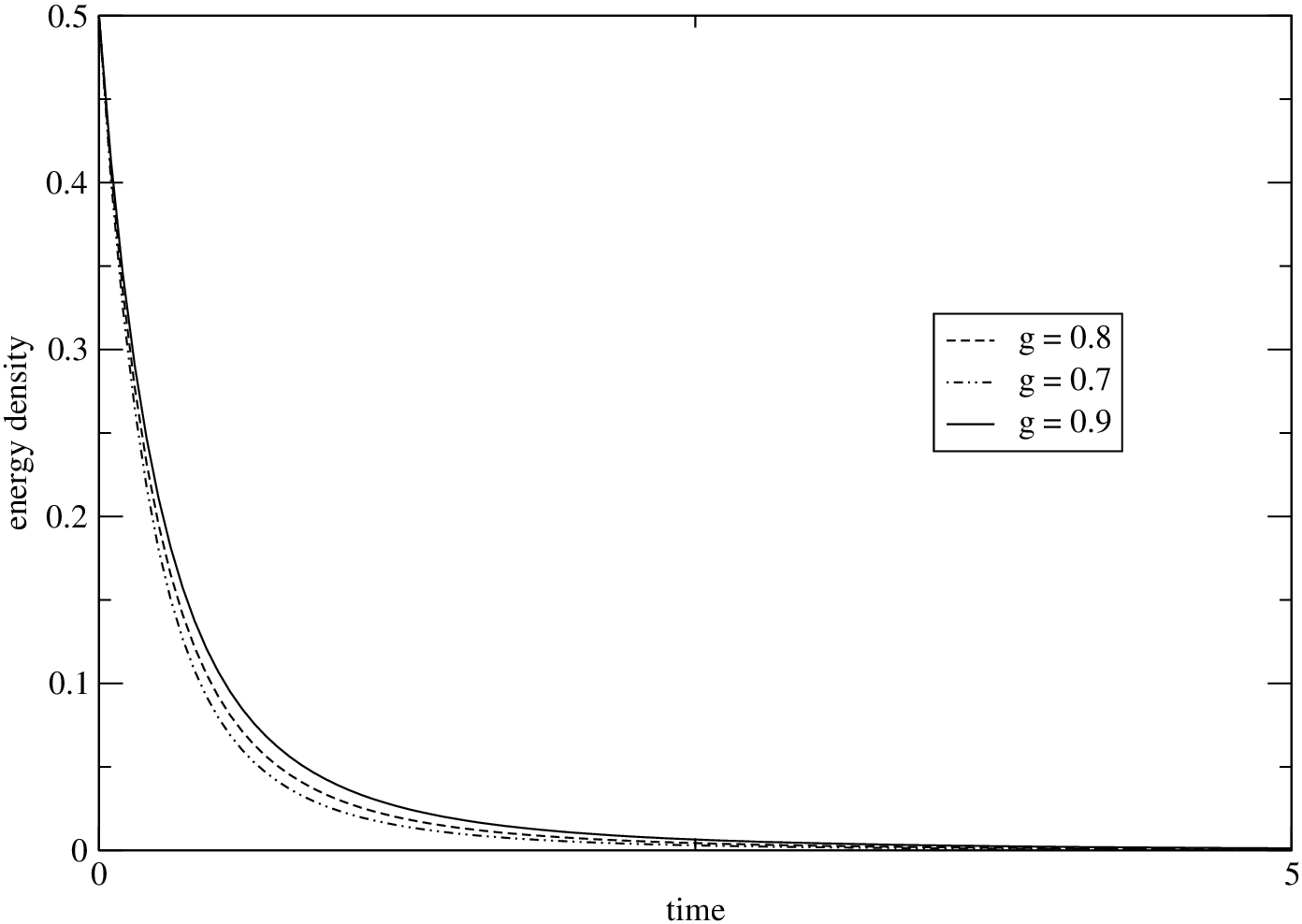}
\caption{evolution in time of vdW  energy density for different values of 
the barotropic parameter in scalar 3D cosmology}
\end{center}
\end{figure}

\noindent The use of this equation of 
state in a
cosmological context was proposed by \cite{Capo1} explored in \cite{Kre} 
and tested in 2D models in \cite{Deve}. 
In our case we have as the dynamical equations

\begin{equation}
2{\ddot a\over a}=-\kappa \left(\rho-{2p}\right)
-({{\dot a}\over{a}})^2+{\lambda},\,\,\, 
\dot\rho+2{\dot a \over a}\left(\rho+p\right)=0\, ,
\end{equation}
\noindent coupled to  the vdW equation

\begin{equation}
p={{g\rho}\over{1-\alpha \rho}}\, .
\end{equation}

\noindent 
This vdW equation without the quadratic term in $\rho$  
was used by Christmann et al. 
(see ref.\cite{Deve}) to investigate 2D Robertson-Walker cosmologies. We are 
taking as initial (normalized) 
values $\rho _0 = 0.7$, $a(0) = 1$,
$\dot a(0) = 1$ and typical values for the vdW parameters $g = 0.9$, $\alpha
= 0.5$.
The one-constituent cosmological fluid ruled by the vdW equation can simulate 
the behavior of a young universe leaving an initial accelerated era as it 
is shown in figure (8). The results also show that if we include a 
cosmological constant a three period evolution is possible, with the initial
accelerated era  followed by  decelerated and finally accelerated periods. The
behavior of the scale factor says that the universe is in an ever expanding 
regime, like in all the previous cases. The evolution of the vdW energy density
is in tune with the rate of expansion of the scale factor (see figure (9)). 
As as final remark we confirm that the three-eras regime is possible 
for  particular values of the parameters like those reported above; 
there are other situations  
(like in the case 
when the normalized initial energy density value is lower than 0.5) 
that a permanent decelerated
evolution appears (with the cosmological constant equal to zero), 
showing that a cosmic fluid obeying  the vdW
equation does not imply necessarily  into an 
initially accelerated 3D universe.

\section{ Conclusions }

In this work we have investigated the presence of   accelerated regimes, 
and their transitions to decelerated eras, in  3D
cosmological models. Two-constituents scenarios are enough to simulate a 
young universe leaving the inflationary period (with the inflaton represented 
by the scalar field) or an old universe coming into a dark energy dominated 
period; the effects of the dark energy constituent can be also  simulated by 
the inclusion of the Chaplygin equation of state. 
 The alternative scalar model, with the cosmological fluid ruled by the 
van der Waals equation of state, permits a three-eras evolution when a cosmological 
constant 
term is present.

\end{document}